\def\kms{{\rm km^{-1}s}}

\def\Mpc{\,{\rm Mpc}}

\def\etal{{\frenchspacing\it et al.}}

\def\rms{{\frenchspacing r.m.s.}}
\documentclass[10pt,emulateapj,apj]{emulateapj}
\usepackage{enumerate}
\usepackage{amsmath}
\def\be{\begin{equation}}\def\bea{\begin{eqnarray}}\def\beaa{\begin{eqnarray*}}
  \def\ee{\end{equation}}  \def\eea{\end{eqnarray}}  \def\eeaa{\end{eqnarray*}}

\shorttitle{Largest Structures in the Universe}
\shortauthors{Park et al.}
\begin{document}
\title{The Challenge of the Largest Structures in the Universe to Cosmology}
\author{Changbom Park$^1$, Yun-Young Choi$^2$, Juhan Kim$^3$, 
J. Richard Gott III$^4$, Sungsoo S. Kim$^{2,5}$, \& Kap-Sung Kim$^{2,5}$}
\affil{$^1$ School of Physics, Korea Institute for Advanced Study, Heogiro 85, Seoul 130-722, Korea \\
$^2$Dept. of Astronomy \& Space Science, Kyung Hee University,
Gyeonggi 446-701, Korea; yy.choi@khu.ac.kr\\
$^3$Center for Advanced Computation, Korea Institute for Advanced Study, Heogiro 85, Seoul 130-722, Korea \\
$^4$Department of Astrophysical Sciences, Peyton Hall, Princeton University, Princeton, NJ 08544-1001, USA \\
$^{5}$School of Space Research, Kyung Hee University,
Gyeonggi 446-701, Korea}
\email{yy.choi@khu.ac.kr}

\begin{abstract}
Large galaxy redshift surveys have long been used to constrain cosmological models and 
structure formation scenarios. In particular, the largest structures discovered 
observationally are thought to carry critical information on the amplitude of 
large-scale density fluctuations or homogeneity of the universe, and have often 
challenged the standard cosmological framework. The Sloan Great Wall (SGW) recently 
found in the Sloan Digital Sky Survey (SDSS) region casts doubt on the concordance 
cosmological model with a cosmological constant (i.e. the flat $\Lambda$CDM model). 
Here we show that the existence of the SGW is perfectly consistent with the 
$\Lambda$CDM model, a result that only our very large cosmological $N$-body simulation 
(the Horizon Run 2, HR2) could supply. In addition, we report on the discovery of a 
void complex in the SDSS much larger than the SGW, and show that such size of the 
largest void is also predicted in the $\Lambda$CDM paradigm. 
Our results demonstrate that an 
initially homogeneous isotropic universe with primordial Gaussian random phase density 
fluctuations growing in accordance with the General Relativity, can explain the 
richness and size of the observed large-scale structures in the SDSS. Using the 
HR2 simulation we predict that a future galaxy redshift survey about four times 
deeper or with 3 magnitude fainter limit than the SDSS should reveal a largest structure of bright galaxies about 
twice as big as the SGW.

\end{abstract}

\keywords{cosmology: large-scale structure of the universe -- cosmology: observations -- galaxies: statistics -- methods: numerical -- methods: observational}
\section{Introduction}
The SGW (Gott {\etal} 2005) found in the SDSS (Aihara {\etal} 2011) is
a thick filamentary 
structure of galaxies located at a distance of about $300\Mpc$ from the Earth. 
Its densest part spans about $200\Mpc$, and the whole filament projected 
on a slice appears to be contiguous over the scale of more than $400\Mpc$. 
It is likely to be longer than the survey size since the structure is cut 
by the survey boundaries. The SGW is reminiscent of the CfA Great Wall 
(Geller \& Huchra 1989), that triggered an intense dispute against the 
`Standard' Cold Dark Matter (SCDM) model (White {\etal} 1987) or 
even the whole class of the gravitational instability models of 
structure formation (Park 1990 and references therein). 
The cosmological principle of homogeneity 
and isotropy of the universe was also doubted since structures as big
as the survey size were always found as the survey size was increased.

The skepticism was relieved when Park (1990) demonstrated that large-scale
structures (LSS's) with size up to $200\Mpc$ can appear in the standard CDM
cosmological model in surveys like the CfA survey. About 20 years later the
SGW revived the doubt on the cosmological principle and the current standard
model ($\Lambda$CDM model) since it is even larger than the CfA Wall 
and comparable with the survey size again
(Sheth \& Diaferio 2011).

In this paper we ask whether or not the existence of the LSS's observed in the 
SDSS is consistent
%; York {\etal} 2000; Blanton {\etal} 2003; 
%Fukugita {\etal} 1996; Gunn {\etal} 1998, 2006; 
%Hogg {\etal} 2001; Ivezic {\etal} 2004; Lupton {\etal} 2001; 
%Pier {\etal} 2003; Smith {\etal} 2002; Stoughton {\etal} 2002;
%Tucker {\etal} 2006
with the $\Lambda$CDM model that adopts a homogeneous isotropic universe 
with primordial Gaussian random phase density fluctuations as predicted 
by inflation and a reasonably successful galaxy assignment scheme. 
The comparison of real superclusters with simulated ones has been made in
several previous studies (Einasto {\etal} 2006; Einasto {\etal} 2007b; 
Einasto {\etal} 2007c; Araya-Melo {\etal} 2009).
We also make predictions on the properties of the LSS's to be observed in the 
future deeper surveys.
Hubble constant of 72 km s$^{-1}\Mpc^{-1}$ is used in this paper.

%%%%%%%%%%%%%%%%%%%%%%%%%%%%%%%%%%%%%%%%%%%%%%%%%%%
%%%%%%%%%%%%%%%%%%%%%%%%%%%%%%%%%%%%%%%%%%%%%%%%%%%
\section{The SDSS Sample}
To identify the LSS's in the observational sample we use the SDSS Main galaxy
sample (Strauss {\etal} 2002), 
which is currently the largest three-dimensional sample of galaxies with 
a high sampling density. A volume-limited subsample of $116,877$ 
galaxies with absolute $r$-band magnitude brighter than $-21.6$ is 
generated from the KIAS value-added catalog (Choi {\etal} 2010a)
, which supplements the bright galaxies missing in the SDSS sample.
The magnitude limit is about $0.6$ magnitude brighter than the critical 
magnitude $M_\ast$ of the SDSS Main galaxy sample (Choi {\etal} 2007), 
and corresponds to the sample depth of $689\Mpc$. 
The mean separation between galaxies is $12.5\Mpc$.
The sample is large enough to reduce the cosmic variance for the number of 
SGW-like structures, and yet the galaxy number density is high enough to 
trace major LSS's. We calculate the comoving distances $R$ of galaxies using the WMAP 
5 year cosmological parameters (Komatsu {\etal} 2009), 
and the Cartesian coordinates are calculated as in Park {\etal} (2007),
\be
x = -R {\rm sin} \lambda, 
y=R {\rm cos} \lambda {\rm cos} \eta, 
z= R {\rm cos} \lambda {\rm sin} \eta,
\ee
where $\lambda$ and $\eta$ are the SDSS survey coordinates.

%%%%%%%%%%%%%%%%%%%%%%%%%%%%%%%%%%%%%%%%%%%%%%%%%%%%%%%%%%%%%%%%%%%%%%%%%%
%%%%%%%%%%%%%%%%%%%%%%%%%%%%%%%%%%%%%%%%%%%%%%%%%%%%%%%%%%%%%%%%%%%%%%%%%%
\section{Identification of the Large-Scale Structures}
Superclusters of galaxies have been identified by many previous studies, which
typically use the smoothed luminosity density of galaxies and apply a threshold
level to define structures (Basilakos, Plionis, \& Rowan-Robinson 2001; 
Einasto {\etal} 2007a; Luparello {\etal} 2011; Liivamagi {\etal} 2012).
Here we adopt the Friend-of-Friend (FoF)
algorithm to identify the high-density LSS's by connecting close galaxies because
the results  
convey the visual impression well and the method
uses effectively only one free parameter.

Before we search for structures in a sample of galaxies, we apply the 
FoF algorithm with the linking length of $3000\,\kms$ to the sample 
to find massive groups. 
The dispersion of each group along the line of sight is forced to be equal 
to that across the line of sight if the former is larger than the latter. 
We then search for LSS's by connecting galaxies separated by less than 
the connection length $d_c$.
A very small $d_c$ results in no LSS's and a very large $d_c$ gives just one 
LSS connecting all galaxies. We choose to use the critical linking length 
that results in the maximum number of structures (Basilakos 2003).
In the case of our volume-limited sample we find the critical linking 
length of $d_c=7.78\Mpc$ when the minimum number of member galaxies 
is set to $20$.

\begin{figure*}
\epsscale{0.93}
\plotone{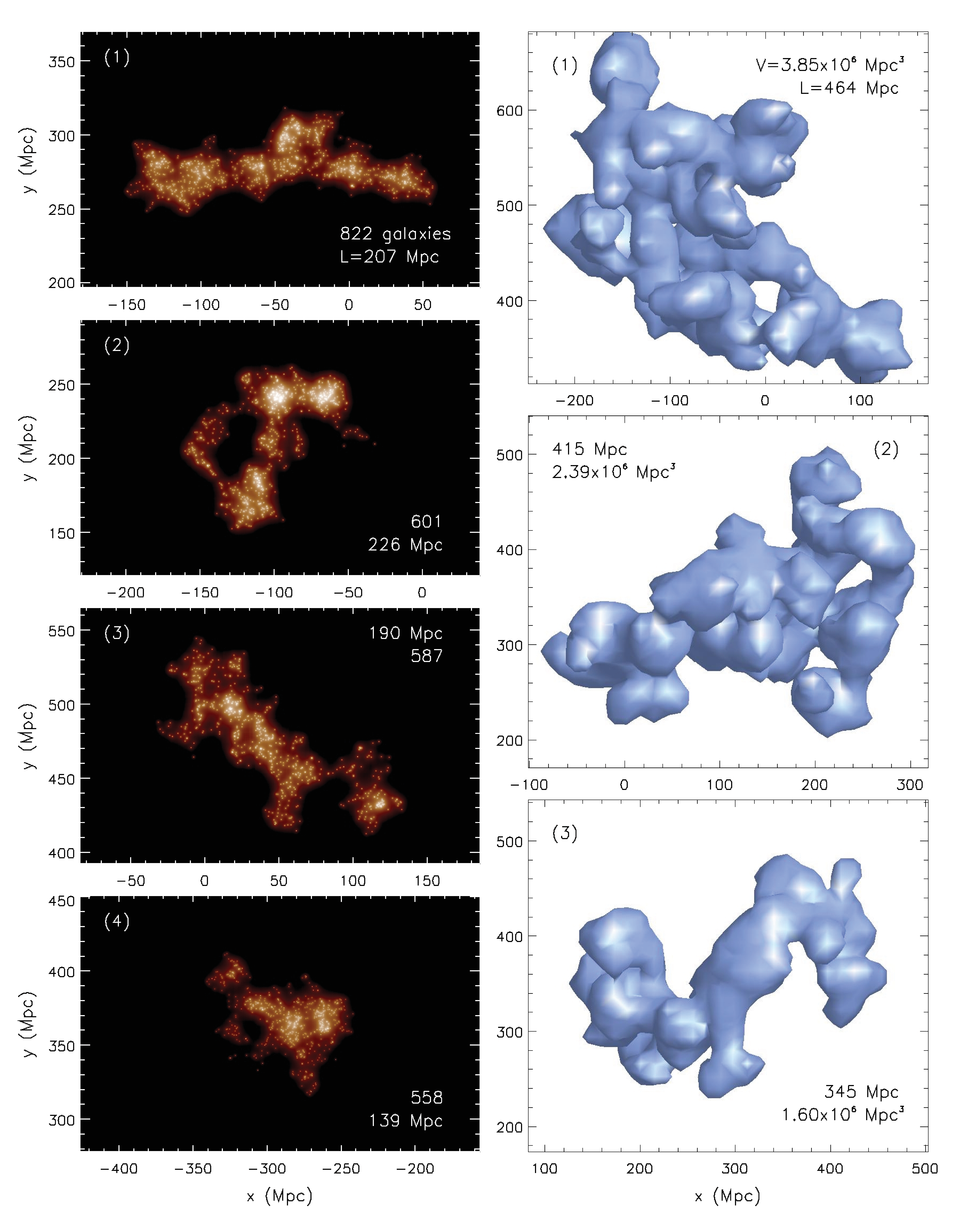}
\caption{
({\it Left}) The four richest high-density large-scale structures found by the 
Friend-of-Friend method with the linking length of $7.78\Mpc$.
Galaxies belonging to each structure are projected on to the $x$-$y$ plane of
the SDSS survey coordinates (Choi {\it et al}. 2010b). 
$L$ is the maximum extent of each structure.
({\it Right}) Three largest volume low-density large-scale structures 
(void complexes). The total volume $V$ is calculated by expanding 
$10.4\Mpc$ to all directions from the core region shown in the plot to take 
into account the boundary regions with the high-density structures.
}
\label{fig:obslss}
\end{figure*}

The left column of Fig.~\ref{fig:obslss} shows the four richest high-density 
LSS's found in the SDSS sample using the FoF method.
The richest one among all $873$ structures discovered within our sample is
a dense part of the SGW containing 822 bright galaxies (the top left panel).
However, the largest one is the second richest structure with the maximum 
extent of $226\Mpc$ containing 601 galaxies. 
The part of the SGW identified by our FoF method is $207\Mpc$ long. 
The richest structures are typically filamentary, and filaments are often looping. 
Also occasionally found are superclusters attaching a few spreading filaments. 
The relatively poor structures are usually single superclusters or doubles 
connected by a thin filament. 
Each high-density LSSs contains a few superclusters found by several 
previous studies (Einasto et al. 2011a; Einasto et al. 2011b; Liivamagi et al. 2012).

The low-density LSS's are found in a similar fashion. We tessellate the SDSS 
region with cubic pixels, and then mark the void core pixels from which there 
is no or only one galaxy within a radius of $r_{\rm ball}$. Note that the 
high-density structures are regions where a ball with a radius of $d_c$
contains two or more galaxies.
Therefore, our definition for void core pixels is consistent with and 
complementary to the definition of the high-density structures.
To identify voids these void pixels contacting their faces are grouped together by
using the FoF algorithm.
We then expand each void out to the distance of $r_{\rm ball}$ - $d_c$ 
to account for the buffer region with high-density structures, 
where $d_c$ is the linking length 
used to identify the high-density structures.
We adopt the critical $r_{\rm ball}=18.14\Mpc$ that results in the maximum 
number of independent voids with volume of the core pixel region 
larger than $(15\Mpc)^3$.

On the right of Fig.~\ref{fig:obslss} the three largest-volume void complexes
found in our SDSS sample are shown. The largest one among the 385 voids found is
located just behind and above (in $y$ and $z$ directions) 
the SGW (the top right panel, see also Einasto {\etal} 2011c, 2012
for a large void at similar location). Its total volume is 
as big as $(157\Mpc)^3$ and its maximum extent is $464\Mpc$, more than 
twice larger than the dense part of the SGW.
The largest few void complexes have very complicated topology showing many 
voids multiply connected by tunnels. The relatively smaller low-density 
structures are usually single voids.

%%%%%%%%%%%%%%%%%%%%%%%%%%%%%%%%%%%%%%%%%%%%%%%%%%%
%%%%%%%%%%%%%%%%%%%%%%%%%%%%%%%%%%%%%%%%%%%%%%%%%%%
\section{$N$-body Cosmological Simulation}
Our main concern is to know how often these largest structures are expected 
to be observed in the currently popular model of the universe.
Taking one more step further we ask if the whole distribution of 
richness or size of the observed LSS's is consistent with the 
prediction of the standard model.

We use a very large cosmological N-body simulation, Horizon Run 2 (HR2)
(Kim {\etal} 2011), for the comparison.
The simulation evolved $6000^3$ particles in a box with a side
length of $10\,$Gpc to calculate the gravitational evolution of primordial
density fluctuations generated in accordance with a $\Lambda$CDM model
(Komatsu {\etal} 2009).
The matter, baryon, and cosmological constant density parameters are set
to $0.26$, $0.044$, and $0.74$, respectively (see Kim {\etal} 2011).
The minimum mass of dark matter subhalos identified in the simulation is
$5.2\times10^{12}$M$_{\odot}$, and the mean subhalo separation is $12.5\Mpc$,
equal to that of our SDSS galaxy sample.
We assume that each dark matter subhalo above the minimum mass contains
one galaxy. This subhalo-galaxy one-to-one correspondence (sHGC) model
has been proved to work well in terms of one-point function and its local
density dependence (Kim {\etal} 2008), two-point function (Kim {\etal} 2009),
and also topology (Gott {\etal} 2009; Choi {\etal} 2010b).
Using this galaxy assignment scheme and the idea of abundance matching
we assume that the subhalos with mass above $5.2\times10^{12}$M$_{\odot}$
compare with our SDSS galaxies brighter than $M_r=-21.6$.

The total volume of our SDSS sample is $(615\Mpc)^3$ effectively.
Our Horizon Run 2 simulation, about 16 times larger in linear size,
is for the first time large enough to capture the large scale power
actually present in the standard model of cosmology and at the same time
has a mass resolution high enough to simulate the SDSS Main Galaxy Sample.
This uniqueness of the simulation enables us
to estimate the statistical likelihood of the LSS's found in the observations.

%%%%%%%%%%%%%%%%%%%%%%%%%%%%%%%%%%%%%%%%%%%%%%%%%%%%%%%%%%%%%%%%%%%%%%%%%%
%%%%%%%%%%%%%%%%%%%%%%%%%%%%%%%%%%%%%%%%%%%%%%%%%%%%%%%%%%%%%%%%%%%%%%%%%%

%%%%%%%%%%%%%%%%%%%%%%%%%%%%%%%%%%%%%%%%%%%%%%%%%%%%%%%%%%%%%%%%%%%%%%%%%%
%%%%%%%%%%%%%%%%%%%%%%%%%%%%%%%%%%%%%%%%%%%%%%%%%%%%%%%%%%%%%%%%%%%%%%%%%%
\section{Comparison between the observations and the $\Lambda$CDM model}
\begin{figure*}
\epsscale{0.93}
\plotone{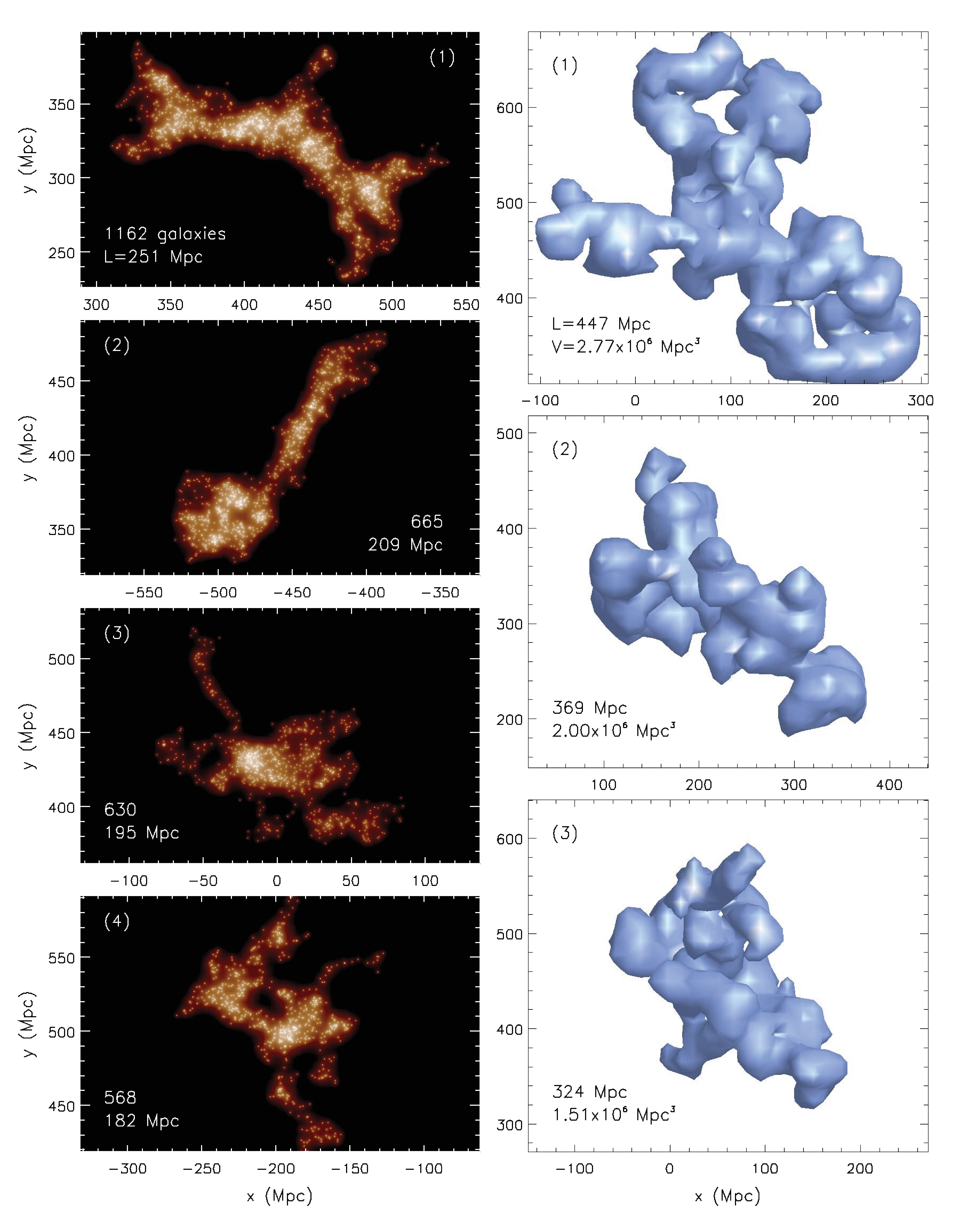}
\caption{
The four largest typical high-density LSS's ({\it left})
and three largest typical low-density structures ({\it right}) selected
from the 200 mock samples.
The high-density LSS's show superclusters and filaments connected quite
similarly to the observed ones. Likewise, the low-density structures
show topology of voids very much like the observed void complexes.
}
\label{fig:mocklss}
\end{figure*}

We make 200 SDSS-like surveys of the `galaxies' in the simulation at the 
present epoch, and analyze the mock survey samples in exactly the 
same way the observational data is analyzed.

For each mock sample the fingers-of-god are identified and compressed, and the 
critical linking length giving the maximum number of LSS's is found.
We find that the mean and standard deviation of the linking lengths from the 
200 mock SDSS samples are $d_c = 7.71\pm 0.18\Mpc$, quite close to 
that of the observational sample.
Fig.~\ref{fig:mocklss} shows the four largest typical high-density LSS's and
three largest typical low-density structures selected from the 200 mock samples.
For example, the largest one in the figure is the structure having the 
approximately median maximum extent and also the median richness
(or volume in the case of low-density structures)
among the 200 largest structures found in the 200 mock samples. 
And the second largest one is the median structure selected from the 
200 second largest ones.

%The LSS’s found in the mock surveys are quite similar to the observations as also indicated by the size and richness distribution functions.

Out of the 200 mocks 137 samples contain a high-density structure richer 
than the SGW. We also find that the largest high-density structure is 
longer than the SGW identified in the same way in 155 cases. Therefore, 
we conclude that structures like the SGW can be easily found in surveys 
like the SDSS in the $\Lambda$CDM universe even though the LSS’s grew from 
primordial Gaussian fluctuations in a homogeneous isotropic background. 
On the other hand, none of the mock samples has the 6-th richest or
largest structure richer or larger than the SGW. This means that even
though the SGW-like structures can be found quite often in a SDSS-like
survey, such large structures are always one of the top six richest and
largest structures. The SGW is indeed a rare object, and was able to be
found because of the large survey volume of the SDSS.

Our conclusion is opposite to that of Sheth \& Diaferio (2011), who used the extreme
value statistics to estimate the likelihood of finding a SGW-like object 
in the SDSS.
They claimed that the existence of the SGW is a $4\sigma$ event in the flat 
$\Lambda$CDM universe with the {\rms} amplitude of density fluctuation 
in a $11.1\Mpc$ radius sphere of $\sigma_8=0.8$, and is difficult to 
reconcile with the model.

\begin{figure}
\epsscale{1.}
\plotone{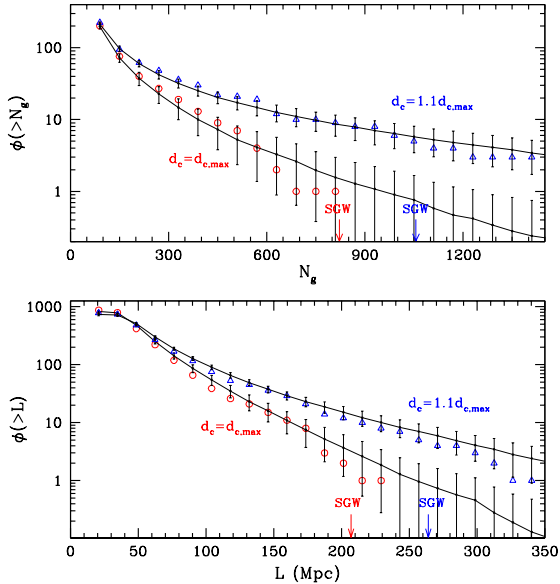}
\caption{
Richness ({\it upper}) and size distributions ({\it bottom}) of high-density 
large-scale structures in the observations and simulations. 
Cumulative distribution functions of the richness and size of the 
high-density LSS's identified in the SDSS sample (circles). 
The richness $N_g$ is the number of member galaxies belonging to each LSS. 
The triangles are the distribution functions when the linking length is 
increased by 10\%. 
The corresponding solid lines and the error bars attached are 
the mean and the standard deviations obtained from 200 mock surveys 
in the Horizon Run 2 simulation. 
The red arrows on the left are the locations of the SGW when the linking 
length $d_c$ is set to the critical value while the right ones are 
when $d_c$ is 10\% larger. 
Note the SGW is the 5th richest and largest structure for this larger 
linking length.
}

\label{fig:hdendist}
\end{figure}

To further inspect the consistency between the observations and the 
$\Lambda$CDM model we calculate the distribution functions of the 
richness and size of the LSS's. Fig.~\ref{fig:hdendist} shows the 
distributions of the number of structures with member galaxies more 
than $N_g$ (open circles in the top panel) and with a maximum extent larger 
than $L$ (open circles in the bottom panel). 
The $y$-axis $\Phi$ is the number of structures per unit SDSS volume. 
The mean and standard deviation of the cumulative histograms from 200 mock 
surveys are shown as lower solid lines and error bars. It can be seen 
that the observed richness and size distributions agree astonishingly 
well with the simulation.

Our $\Lambda$CDM simulation tells us that on average the richest 
high-density LSS in the flat $\Lambda$CDM universe in a SDSS volume 
is expected to contain 957 galaxies brighter than $M_r = -21.6$ when the 
linking length is $7.78\Mpc$, and that the typical size of the 
largest structure is $255\Mpc$. 
These values compare with 822 galaxies and $226\Mpc$ for the SDSS sample. 
Therefore, the largest structures in the observations are actually a 
little smaller than the $\Lambda$CDM expectations both in the richness and size
.
The triangles and upper solid lines are the cumulative histograms 
for the observation and the simulation when the linking length is increased 
by 10\% above the critical value. 
The plots demonstrate that the agreement does not depend on the 
choice of the linking length.

Springel {\etal} (2006) claimed that a SGW-like object was found in their 
Millennium Simulation to support their view that LSS does not provide the 
strongest challenge to $\Lambda$CDM.
However, considering the fact that the matter fluctuation power spectrum 
of their simulation deviated largely from the $\Lambda$CDM theory near 
the simulation box scales due to incorrect normalization and 
statistical fluctuations (Springel {\etal} 2005) and 
also having only one simulation whose volume is roughly equal to SDSS, 
it would be difficult to draw such a conclusion on the prevalence of a 
SGW-like structure in the SDSS-like surveys in the $\Lambda$CDM universe. 
Our HR2 is about $4,300$ times larger than the volume of the SDSS sample 
with the fundamental mode more than 15 larger than the depth of SDSS, and 
accurate statistical comparisons as presented here are possible.

\begin{figure}
\epsscale{1.}
\plotone{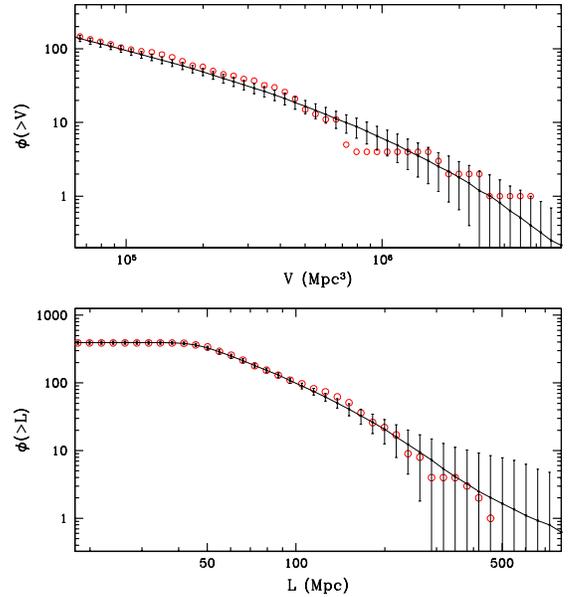}
\caption{
Volume ({\it upper}) and size distributions ({\it bottom}) of voids 
in the observations and simulations. 
Cumulative distribution functions of the volume and size of the 
void complexes identified from the SDSS sample (circles). 
Two hundred mock surveys in the Horizon Run 2 simulation are used to 
calculate the mean (solid lines) and standard deviations (error bars).
}

\label{fig:ldendist}
\end{figure}

The impressive agreement between the observations and the $\Lambda$CDM model 
is also found for the volume and size of void complexes. 
Fig.~\ref{fig:ldendist} shows the number of the void complexes in the SDSS 
sample with volume larger than $V$ (open circles in the upper panel) and 
that with the maximum extent larger than $L$ (open circles in the bottom panel). 
The cumulative distribution functions agree very well with the mean of 
the 200 mock surveys in the HR2 (solid lines with error bars).

To find how the size of the largest LSS scales as the survey size 
increases we made 27 non-overlapping mock surveys in the HR2 simulation
having the SDSS angular mask but 
with the outer boundary located at $2,767\Mpc$ or redshift $0.8$. 
These surveys are about 4 times deeper or 3 magnitude fainter than the SDSS. 
We use these mock surveys to correctly account for the survey boundary
effects on the scaling of the LSS.
We find that the largest LSS typically has the mean number of galaxies 
of $2,480$ and the maximum extent of $430\Mpc$. Therefore, the maximum 
richness and size will increase by a factor of only about 3 and 1.7,
respectively, and 
the universe will look more homogeneous over the scale of the survey size.

%%%%%%%%%%%%%%%%%%%%%%%%%%%%%%%%%%%%%%%%%%%%%%%%%%%%%%%%%%%%%%%%%%%%%%%%%%%%%%%%%%%%%%%%%%%%%%%%%%%%%%%%%%%%%%%%%%%%%%%%%%%%%%%%%%%%%%%%%%%%%%%%%%%%%%%%%%%%%%%%%
\section{Conclusions}

We identify high-density and low-density LSS's in the SDSS to test whether or not
the current standard $\Lambda$CDM model of the universe can explain 
the observed structures of bright galaxies (Choi {\etal} 2010b).
The LSS's used in this comparison are those when the
characteristic connection lengths result in the maximum number of structures.
It is found that the richest high-density structure is the dense part of the 
SGW, which is also the second largest structure.
The low-density LSS's are typically much larger than the 
high-density counter parts, and the largest one is found to be $464\Mpc$ long.

The Horizon Run 2 simulation of the $\Lambda$CDM universe is used to make 
a set of mock SDSS surveys. Galaxies are assigned to dark matter 
subhalos assuming that each halo contains one galaxy and adopting the abundance 
matching (Kim {\etal} 2008). 
LSS's are identified in the exactly same way that the SDSS data is analyzed.
We find that the structures with richness and size similar to the SGW 
are usually one of the richest and/or the largest structures in the mock samples. 
To estimate the statistical significance of the largest observed LSS's
and to check the consistency of the properties of the observed LSS's
with those of the structures found in $\Lambda$CDM we measure the
distribution functions of richness and maximum extent of LSS's
identified in the SDSS and the mock samples.
It is found that the observed distribution functions agree with those of
the simulation astonishingly well.
We conclude that both observed high-density and low-density LSS's have the
richness/volume and size distributions consistent with the $\Lambda$CDM universe.
This agreement between the observations and the theoretical predictions should be
considered as one of the great successes of the $\Lambda$CDM cosmological model.

Einasto et al. (2006, 2007b, 2007c) compared properties of 2dFGRS and SDSS 
superclusters with those of superclusters identified from the Millennium Run 
mock galaxy catalogue. They showed that the geometric properties of real 
superclusters such as the size, the degree of asymmetry and compactness, and 
the mass of the richest superclusters are similar to those of simulated ones 
(see also Araya-Melo et al. 2009). It was found that the fraction of such extremely 
massive and richest superclusters is too small in the simulated samples 
when compared to the observed samples, and the morphology of the richest 
supercluster in the SGW is not recovered in simulation (see also Einasto et al. 2011b). 
Richness and luminosity function of LSS’s depend sensitively on three things: 
first, the initial conditions such as the primordial power spectrum, 
second, the galaxy (and their luminosity) assignment scheme, and 
third, the LSS identification method. Therefore, for a fair comparison 
between observations and simulations, it is very important to use the 
same mass objects and to identify the LSS's using the same criteria. 
Whether or not the largest LSS's of the $\Lambda$CDM universe have properties 
different from the observed ones remains to be studied further.

We note in this study that the properties of LSS's depend
sensitively on the initial power spectrum and the growth of structures and 
can be a powerful tool to discriminate among cosmological models and
galaxy formation scenarios. 
We plan to further explore the usefulness of using the LSS properties in 
cosmology in the future studies.

%%%%%%%%%%%%%%%%%%%%%%%%%%%%%%%%%%%%%%%%%%%%%%%%%%%%%%%%%%%%%%%%%%%%%%%%%%%%%%%%%%%%%%%%%%%%%%%%%%%%%%%%%%%%%%%%%%%%%%%%%%%%%%%%%%%%%%%%%%%%%%%%%%%%%%%%%%%%%%%%%
\begin{acknowledgements}
This work was supported by a grant from the Kyung
Hee University in 2011 (KHU-20100179). We thank 
Graziano Rossi for helpful comments and 
Korea Institute for Advanced Study for providing computing
resources (KIAS Center for Advanced Computation
Linux Cluster System) for this work.

Funding for the SDSS and SDSS-II has been provided by
the Alfred P. Sloan Foundation, the Participating Institutions,
the National Science Foundation, the U.S. Department of
Energy, the National Aeronautics and Space Administration,
the Japanese Monbukagakusho, the Max Planck Society, and
the Higher Education Funding Council for England. The SDSS
Web site is http://www.sdss.org/.

%The SDSS is managed by the Astrophysical Research Consortium
%for the Participating Institutions. The Participating
%Institutions are the American Museum of Natural History,
%Astrophysical Institute Potsdam, University of Basel, Cambridge
%University, Case Western Reserve University, University
%of Chicago, Drexel University, Fermilab, the Institute
%for Advanced Study, the Japan Participation Group, Johns
%Hopkins University, the Joint Institute for Nuclear Astrophysics,
%the Kavli Institute for Particle Astrophysics and Cosmology,
%the Korean Scientist Group, the Chinese Academy of Sciences
%(LAMOST), Los Alamos National Laboratory, the Max-Planck-Institute 
%for Astronomy (MPIA), the Max-Planck-Institute for
%Astrophysics (MPA), New Mexico State University, Ohio State
%University, University of Pittsburgh, University of Portsmouth,
%Princeton University, the United States Naval Observatory, and
%the University of Washington.
\end{acknowledgements}
%%%%%%%%%%%%%%%%%%%%%%%%%%%%%%%%%%%%%%%%%%%%%%%%%%%%%%%%%%%%%%%%%%%%%%%%%%%%%%%%%%%%%%%%%%%%%%%%%%%%%%%%%%%%%%%%%%%%%%%%%%%%%%%%%%%%%%%%%%%%%%%%%%%%%%%%%%%%%%%%%

% BIBLIOGRAPHY

%%%%%%%%%%%%%%%%%%%%%%%%%%%%%%%%%%%%%%%%%%%%%%%%%%%%%%%%%%%%%%%%%%%%%%%%%%%%%%%%%%%%%%%%%%%%%%%%%%%%%%%%%%%%%%%%%%%%%%%%%%%%%%%%%%%%%%%%%%%%%%%%%%%%%%%%%%%%%%%%%
\end{document}